\documentclass[11pt, letter]{amsart}
\usepackage{amsthm,amsmath,amsxtra,amscd,amssymb,xypic,color,xspace,esint}
\usepackage{latexsym,amsmath,amssymb,mathrsfs}
\usepackage{amsfonts,eucal,amsthm,graphicx,color}

\usepackage{graphicx} 
\usepackage{hyperref}

\hypersetup{
    colorlinks = true,
   linktocpage = true,
    citecolor = {blue}
}

\usepackage{algpseudocode}
\usepackage{subcaption}
\usepackage{svg}

\usepackage{latexsym}
\usepackage{amsmath}
\usepackage{amssymb}
\usepackage{mathrsfs}
\usepackage{amsfonts}
\usepackage{eucal}
\usepackage{amsthm}
\usepackage{graphicx}
\usepackage{color}
\usepackage{youngtab}
\usepackage{slashed}
\usepackage{tikz-cd}

\newcommand {\qe}{\mathfrak{q}}

\newcommand {\CalC} {\mathcal C}

\newcommand {\CalN} {\mathcal N}
\newcommand {\CalO} {\mathcal O}

\newcommand {\BC}   {\mathbb C}

\newcommand {\BR}   {\mathbb R}

\newcommand {\BZ}   {\mathbb Z}

\newcommand {\ve}{\varepsilon}

\newcommand{\bch}{\boldsymbol{\chi}}
\newcommand{\bmu}{\boldsymbol{\mu}}

\newcommand{\ii}{\mathrm{i}}

\newcommand{\beq}{\begin{equation}}
\newcommand{\eeq}{\end{equation}}

\newtheorem{theorem}{Theorem}[section]

\begin{document}

\title[Elliptic Vershik-Kerov formula]{Elliptic analogue of \\
Vershik-Kerov limit shape}
\author{Andrei Grekov, Nikita Nekrasov}
\address{Simons Center for Geometry and Physics$^{n}$\\
Yang Institute for Theoretical Physics$^{g,n}$\\ 
Stony Brook University, Stony Brook NY 11794-3636, USA}

\begin{abstract}
We review the limit shape problem for the Plancherel measure and its generalizations found in supersymmetric gauge theory instanton count. We focus on the measure, interpolating between the Plancherel measure and uniform measure, a $U(1)$ case of ${\CalN}=2^{*}$ gauge theory. We give the formula for its limit shape in terms of elliptic functions, generalizing the trigonometric ``arcsin'' law of Vershik-Kerov and Logan-Schepp\\

Dedicated to the 90th anniversary of Anatoly Moiseevich Vershik, with admiration
\end{abstract}

\maketitle
\section{Introduction}
In a seminal paper \cite{VK} A. Vershik and S. Kerov studied the large $N$ asymptotics of the Plancherel measure on the set of irreducible representations $R_\lambda$ of symmetric group $S(N)$:
\beq
    \mu [\lambda] = \frac{(\text{dim} R_\lambda)^2}{N!}
    \label{eq:Plancherel}
\eeq
To $R_{\lambda}$ one associates Young diagram $\lambda$
\beq
{\lambda} = ({\lambda}_{i})\, , \ {\lambda}_{1} \geq {\lambda}_{2} \geq \ldots \geq {\lambda}_{{\ell}({\lambda})} > 0
\eeq
with 
\beq
| {\lambda} | = N  = {\lambda}_{1} + {\lambda}_{2} + \ldots + {\lambda}_{{\ell}({\lambda})}
\eeq
boxes. The main result of \cite{VK} is that upon rescaling the linear size of $\lambda$ by $\sqrt{N}$ one finds, in the $N \to \infty$ limit, 
a piecewise smooth curve $f(x)$, \emph{the arcsin law}, which is read off a certain rational curve $\Sigma$ through a solution of a Riemann-Hilbert problem.  This limit shape curve determines the large $N$ asymptotics of expectation values of all the moments (cf. \cite{OP})
\beq
{\bf p}_k [{\lambda}] = (1-2^{-k}) {\zeta}(-k) + \sum_{i=1}^{\infty}  \left( {\lambda}_{i} - i  + \frac 12 \right)^{k} -  \left(  - i  + \frac 12 \right)^{k}
\label{eq:pkmom}
\eeq
Random Young diagrams behave in a way, similar to large $N$ random $N \times N$  matrices.

In this small note we will study a one-parametric family of random partition models, motivated by the studies of supersymmetric gauge theories in four dimensions \cite{NO}. The measure \eqref{eq:Plancherel} arises in a limit. To be more precise, the original Vershik-Kerov problem can be equivalently studied in the macrocanonical ensemble, where one sums over all $N$ with 
the weight $\frac{1}{N!} z^{N}$, with the parameter $z$ called fugacuty. Instead of the large $N$ limit one studies the large $z$ limit. It is this ensemble that we generalize below. 

The paper is organized as follows. 
In section $\bf 2$ we rederive the main result of \cite{VK} using the language of $qq$-characters \cite{NP, BPSCFT1}.  In section $\bf 3$ 
we introduce our generalization of the macrocanonical ensemble, the corresponding $qq$-character. In section $\bf 4$ we solve the limit shape problem by using the factorization of the \emph{theta-transform} of the $qq$-character. In section $\bf 5$ we discuss various limits of our solution: comparison to Vershik-Kerov limit shape, as well as the edge behavior. In section $\bf 6$ we discuss future directions, including the \emph{higher times} generalizations of the problem. 

\section{Profiles, limit shapes, the arcsin law}
We will use the results and notations from \cite{NO}. 

The Plancherel measure \eqref{eq:Plancherel} above is a probablity measure on the set of Young diagrams of fixed size $N$, as
\beq
\sum_{\lambda \, , \ |{\lambda}| = N} {\mu}[{\lambda}] = 1
\eeq
In the studies of four dimensional gauge theory \cite{Nekrasov:2002qd} one arrives at the similar measure, but defined on the set of all Young diagrams, the size $N = |{\lambda}|$
being weighted with a fugacity factor
\beq
{\bmu}_{\Lambda, \hbar} ({\lambda}) = \frac{1}{Z} \frac{1}{N!} \left( \frac{\ii\Lambda}{\hbar} \right)^{2N} {\mu} [ {\lambda} ]
\label{eq:macro}
\eeq
where the parameters $\Lambda$ and $\hbar$ have the meaning of the instanton counting parameter and the $SU(2)$-rotation equivariant parameter, respectively. The normalization factor $Z$ is given by:
\begin{equation} \label{partfunct}
    Z = \sum_{\lambda} \Big(\frac{ \ii \Lambda}{\hbar} \Big)^{2|\lambda|} \prod_{\square \in \lambda} \frac{1}{ h_\square^2}
\end{equation}
where $h_\square$ - is a hook length of a box $(i,j)$ in the Young diagram $\lambda$:
\beq
h_{i,j} = {\lambda}_{i} - j + {\lambda}^{t}_{j} - i +1 
\label{eq:hook}
\eeq
Let $x$ be an indeterminate. 
Define the ${\bf Y}(x)$-observable on the set of all Young diagrams by
\beq
{\bf Y}(x) \vert_{\lambda} = x \prod_{\square \in \lambda} \frac{(x- \hbar c_\square)^2 - \hbar^2}{(x- \hbar c_\square)^2}
\label{eq:yobs}
\eeq
where for $\square = (i,j) \in {\lambda}$ its \emph{content} is defined by
\beq
c_{i,j} = i-j 
\eeq
By cancellation of factors in the product it could be also written as:
\begin{multline} \label{Ycancelled1}
{\bf Y}(x) \vert_{\lambda} = x \prod_{i = 1}^{{\ell}_{\lambda}} \frac{x - \hbar i}{ x  + \hbar ( \lambda_i - i) }  \frac{x + \hbar({\lambda}_{i} - i +1 )}{x + \hbar (1-i)} = 
\frac{\prod_{{\square} \in {\partial}_{+}{\lambda}} (x - {\hbar}c_{\square})}{\prod_{{\blacksquare} \in {\partial}_{-}{\lambda}} (x - {\hbar}c_{\blacksquare})}
\end{multline}
where $\lambda_i = 0$ for $i > \ell_{\lambda} = {\lambda}_{1}^{t}$, and $\partial_{+}{\lambda}$ and $\partial_{-}{\lambda}$ are the sets of boxes which can be added  to or removed from $\lambda$, respectively.\\
Obviously, the expectation value
\begin{equation}
    \langle {\bf Y} (x) \rangle: = \sum_{\lambda} \,  {\mu}_{\Lambda, \hbar} [{\lambda} ] \, {\bf Y}(x) \vert_\lambda
\end{equation}
has poles as a function of $x$. 
Define another observable, the \emph{character}
\beq
{\bch}(x)\vert_\lambda = {\bf Y}(x)\vert_\lambda + \frac{\Lambda^2}{{\bf Y}(x)\vert_\lambda}
\label{eq:qchar}
\eeq
It was proven in \cite{BPSCFT1}, that the expectation value
$\langle {\bch}(x) \rangle$
 of the $qq$-character is pole-free. From \eqref{eq:yobs}, \eqref{eq:qchar} we see that:
\begin{equation} \label{asymptY}
    {\bf Y}(x) , {\bch}(x)  \sim x + O\Big(\frac{1}{x^2}\Big)\, , 
    \, \text{as} \,\, x \rightarrow \infty
\end{equation}
hence:
\begin{equation}\label{qqChar}
\langle {\bch}(x) \rangle = x     
\end{equation}
Now we are ready to apply the main result of \cite{NO}. Namely, in the limit $\hbar \rightarrow 0$, the correlation functions defined using the measure \eqref{eq:macro} factorize, 
\beq
\langle {\CalO}_{1} {\CalO}_{2} \rangle = \langle {\CalO}_{1} \rangle \langle {\CalO}_{2} \rangle + o({\hbar}) \, , \ {\hbar} \to 0
\label{eq:factoriz}
\eeq
thus they become evaluations on the limit shape $\lambda^\infty$, defined as a $C^0$ limit $f(x)$ (in fact, the limit is in $C^1$) of the profile function $f_{\lambda}(x)$ (cf. \cite{NO}):
\begin{multline}
f_{\lambda}(x) : = | x | + \\
+ \sum_{i=1}^{\infty} \left( \, |x - {\hbar}({\lambda}_{i} - i +1) | - |x - {\hbar}({\lambda}_{i}-i)| + | x + {\hbar} i| - | x + {\hbar}(i-1)| \, \right) \ .
\label{eq:profile}
\end{multline}
We give an example of profile $f_{\lambda}(x)$ below (\ref{profile}):
\begin{figure}[h]
    \centering
    \includegraphics[scale = 0.4]{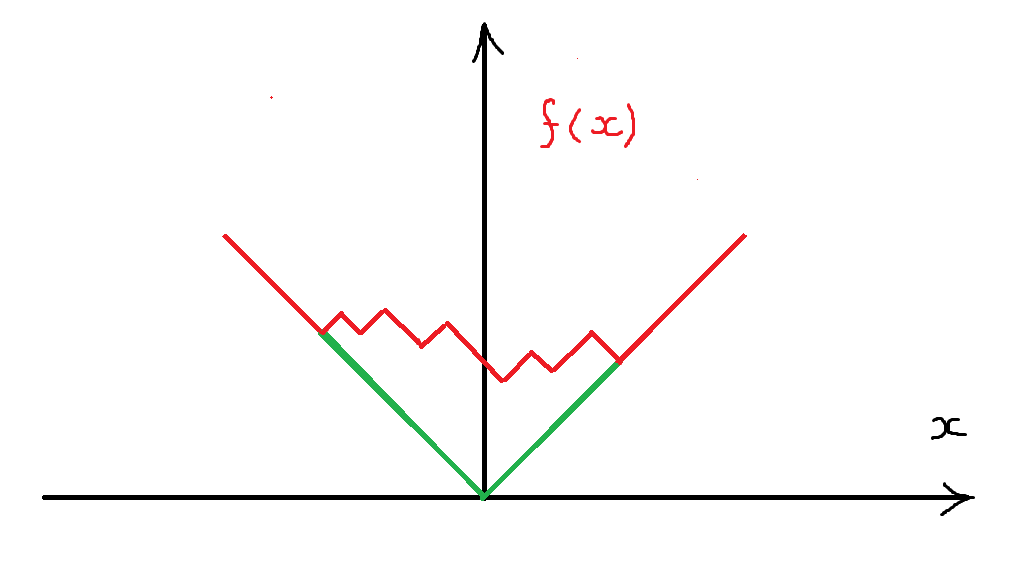}
    \caption{Young diagram profile function}
    \label{profile}
\end{figure}
The ${\bf Y}(x)$-observable expresses through the profile function via \cite{NO}:
\begin{equation} \label{Ylam}
    {\bf Y}(x) \vert_{\lambda} \, =\,   \exp\Big[\frac{1}{2}  \int_\mathbb{R} \log(x-y) f_{\lambda}''(y) dy \Big]
\end{equation}
It follows from the formula (\ref{Ycancelled1}) by direct calculation.
The factorization \eqref{eq:factoriz} is proven using the standard argument \cite{VK}: the measure \eqref{eq:macro}
scales as
\begin{multline}
{\bmu}_{\Lambda, \hbar} ({\lambda} ) \, = \, \left( 1 + O({\hbar}) \right) \, \times  \\
e^{\frac{1}{2 \hbar^2} \, {\rm P.V.} \, \int_{{\BR}^{2}} dx_{1} dx_{2} f_{\lambda}^{''}(x_1) f_{\lambda}^{''}(x_2)  ( x_{1} - x_{2} )^2 \left( {\rm log} \left( \frac{x_{1}  - x_{2}}{\Lambda}  \right) - \frac 32 \right)} 
\label{eq:meashbar}
\end{multline}
while the entropy factor, the number of configurations $\lambda$ whose profile is $C^0$-close to $f(x)$,  grows as 
\beq
\propto e^{c \frac{L_{\lambda}}{\hbar}} 
\eeq
where $L_{\lambda}$ is a length of the boundary of $\lambda$, measured in $\hbar$-units, $L = {\hbar} \left( {\lambda}_{1}^{t} + {\lambda}_{1} \right)$, and $c$ is a constant of order $1$. Thus, in $\hbar \to 0$ limit
\beq
\langle {\bf Y}(x) \rangle \to Y(x)\, , \ \langle{\bf Y}(x)^{-1} \rangle \to \frac{1}{Y(x)}
\eeq
with 
\beq\label{Ylimit}
    Y(x)  \, =\,   \exp\Big[\frac{1}{2}  \int_\mathbb{R} \, {\log}(x-y) \, f''(y) dy \Big]
\eeq
so that \eqref{qqChar}
becomes the equation of the rational curve:
\begin{equation} \label{ratcurve}
    x = Y(x) + \frac{\Lambda^2}{Y(x)}
\end{equation}
Taking the derivative of (\ref{Ylimit}) one gets:
\begin{equation}
    G(x):=  \frac{d}{dx} \log Y(x) = \frac{1}{2} \int_\mathbb{R} \frac{f''(y)}{x-y} dy \, , 
\end{equation}
a function admitting analytic continuation to the complex plane $x \in {\BC}$, with branch cut on the support of $f''(x)$. The jump of $G(x)$ across the cut is equal to:
\begin{equation}
    G(x+ {\ii} 0) - G(x-{\ii} 0) \, =\,  {\ii\pi} f''(x)
\end{equation}
To read off $f''(x)$ we compare the expression above to the solution of (\ref{ratcurve}):
\begin{equation}
    Y(x) = \frac{x}{2} + \frac{1}{2} \sqrt{x^2 - 4 \Lambda^2}
    \label{eq:yplus}
\end{equation}
The Eq. \eqref{ratcurve} has two solutions for $Y(x)$ in terms of $x$. The ``$+$'' branch of the square root in \eqref{eq:yplus} is chosen so as to give the correct large $x$ asymptotics (\ref{asymptY}). Taking the derivative we arrive at:
\begin{equation}
    G(x) = \frac{1}{\sqrt{x^2 - 4 \Lambda^2}}
\end{equation}
It has a branch cut from $-2 \Lambda$ to $2 {\Lambda}$. Calculating the jump across it we arrive at the expression for $f''(x)$ (for $|x| < 2\Lambda$):
\begin{equation}
    f''(x) = \frac{1}{ \pi \Lambda } \frac{1}{\sqrt{1-\big(\frac{x}{2\Lambda}\big)^2  }}
    \label{eq:limitVK}
\end{equation}
The size of the corresponding partition asymptotes to
\beq
N \sim \frac{1}{4\hbar^2} \int_{\BR} f^{''}(x) x^{2} = 
\frac{\Lambda^2}{2\hbar^2}  
\eeq
Integrating \eqref{eq:limitVK} once gives:
\begin{equation} \label{fprime}
    f'(x) = \frac{2}{\pi} \arcsin \frac{x}{2{\Lambda}}
\end{equation}
and integrating twice we are proving the following theorem:
\begin{theorem} (Vershik-Kerov-Logan-Schepp'77, \cite{VK, LS})
    The limit shape of the distribution \eqref{eq:Plancherel} on the set of Young diagrams of size $N$
    as $\hbar \sim \frac{\Lambda}{\sqrt{2N}} \rightarrow 0$ is described by the following profile:
    \begin{equation}
        f_{VK}(x) = \begin{cases}
            \frac{2}{\pi} \Big( x \arcsin \frac{x}{2 {\Lambda}} + \sqrt{4 \Lambda^2 - x^2} \Big), \,\, |x| \leq 2 {\Lambda} \\
            |x|, \,\, |x| \geq 2 {\Lambda}
        \end{cases}
        \label{eq:VKlim}
    \end{equation}
    representing $\hbar$-rescaled piece-wise linear boundary of $\lambda$.
\end{theorem}
{\bf Remark.} Mathematicians usually present \eqref{eq:VKlim}
with $\Lambda = 1$. Keeping $\Lambda$ as a parameter
is motivated by generalizations involving, e.g. several random Young diagrams \cite{Nekrasov:2002qd}. 

\section{Elliptic generalization of Vershik-Kerov model}
In this section we are introducing a one parameter deformation of \eqref{eq:macro}. It arises in the studies of mass deformed maximally supersymmetric gauge theory, the so-called ${\CalN}=2^{*}$ theory \cite{Nekrasov:2002qd}:
\begin{equation} \label{partfunctmassive}
    {\bmu}_{m, {\qe}, \hbar} [ {\lambda} ] \, = \, \frac{{\qe}^{|\lambda|}}{Z_{2^{*}} (m, {\qe}, \hbar)}  \prod_{\square \in \lambda} \left( 1 -  \left( \frac{m}{\hbar h_{\square}} \right)^2 \right)\end{equation}
with the normalization partition function $Z_{2^{*}}$ defined so that 
$\sum_{\lambda} {\bmu}_{m, {\qe}, \hbar} [ {\lambda} ] = 1$\footnote{In \cite{NO,BPSCFT1} an explicit formula for $Z_{2^{*}}$ can be found, but we don't need it here}. 
The fugacity    
\beq
{\qe} = e^{2 \pi i \tau }    
\eeq
is usually written in terms of the modular parameter 
\beq
\tau = \frac{\vartheta}{2\pi} + \frac{4\pi\ii}{g^2}
\eeq
of elliptic curve underlying the microscopic ${\CalN}=4$ theory \cite{DW}, in agreement with the Montonen-Olive S-duality conjecture \cite{VW}. Mathematically the measure makes sense for any complex values of $m$ and $\mathfrak{q}$ such that $|\mathfrak{q}|<1$, however in order for it to be a positive definite distribution on the set of all Young diagrams we restrict $\mathfrak{q} \in \mathbb{R}$, and  $m \in \ii\mathbb{R}$. \\
The ${\bf Y}(x)$-observable is defined as before by \eqref{eq:yobs}, while the $qq$-character is much more involved (cf. Eq. (153) in the arxiv version of \cite{BPSCFT1}, see also \cite{NP}, where the $\hbar\to 0$ limit was analyzed):
\begin{equation}
    {\bch}(x) = \sum_{\nu} {\bmu}_{{\hbar}, {\qe}, m} [ {\nu} ]\,   \frac{\prod_{\square \in \partial_+\nu }   {\bf Y}(x + m c_\square)}{\prod_{\square \in \partial_-\nu }   {\bf Y}(x + m c_\square)} \ . 
    \label{eq:A0char}
\end{equation}
In \eqref{eq:A0char} the sum is taken over the set of auxiliary Young diagrams $\nu$, not to be confused with the diagrams $\lambda$ of the original ensemble \eqref{partfunctmassive}. Note that the roles of $m$ and $\hbar$ in \eqref{eq:A0char} are switched compared to 
\eqref{partfunctmassive}. 

The main theorem of \cite{BPSCFT1} implies that the expectation value:
\begin{equation}
    \langle {\bch}(x) \rangle_{2^{*}}: =  \sum_{\lambda} \, {\bmu}_{m,{\qe}, {\hbar}}[{\lambda}]
\, {\bch}(x) \vert_\lambda \, \end{equation}
has no poles in $x$, behaves as $x$ for large $x$, therefore it is equal to $x$: 
\beq
\langle {\bch}(x) \rangle_{2^{*}} = x \ .
\label{eq:vevA0}\eeq

\section{Solving for limit shape in the elliptic case}

\begin{figure}[h]
    \centering
    \includegraphics[scale = 0.6]{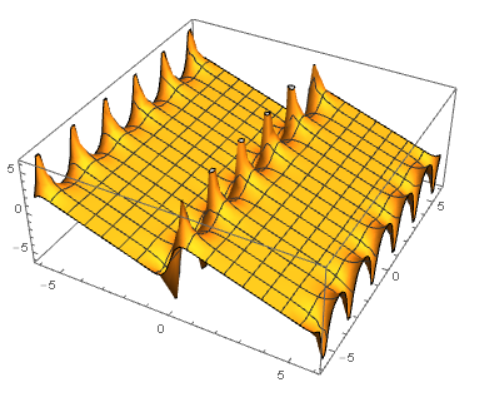}
    \caption{Re$[x(z)]$}
    \label{E1}
\end{figure}

\begin{figure}[h]
    \centering
    \includegraphics[scale = 0.5]{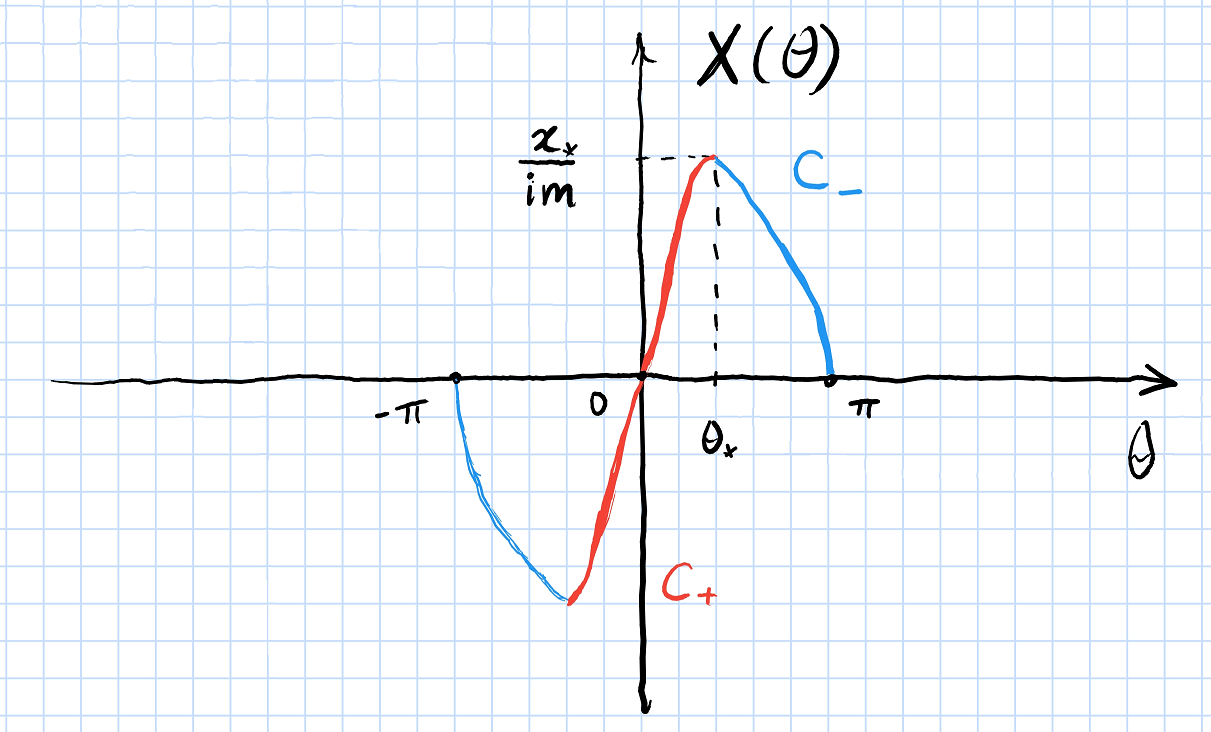}
    \caption{$X(\theta)$ with colored sides of the cut}
    \label{mapgraph}
\end{figure}

\begin{figure}[h]
    \centering
    \includegraphics[scale = 0.6]{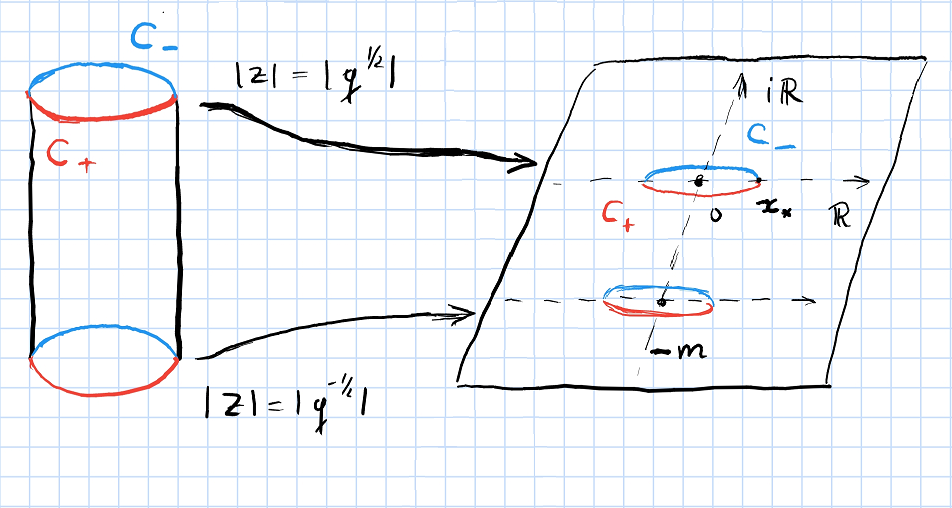}
    \caption{Domain and image of $x(z)$}
    \label{mappict}
\end{figure}

In the limit $\hbar \rightarrow 0$, the same arguments as in the previous section show the expectation values of ${\bf Y}(x)$ tend to evaluations on the limit shape $\lambda^\infty_{2^{*}}$, 
\begin{multline}
\langle {\bf Y}(x) \rangle_{2^{*}} \xrightarrow[\hbar \to 0]{} \ Y(x)\, , \\
\langle {\bch}(x) \rangle_{2^{*}}\xrightarrow[\hbar \to 0]{} \, 
{\phi}({\qe}) \sum_{\nu} \mathfrak{q}^{|\nu|}   \frac{\prod_{\square \in \partial_+\nu }   Y(x + m c_\square)}{\prod_{\square \in \partial_-\nu }   Y(x + m c_\square)}
\label{eq:charA0}
\end{multline}
where
\beq
{\phi}({\qe})=\prod_{n=1}^{\infty} (1-{\qe}^{n}) \ .
\eeq
Comparing \eqref{eq:charA0} and \eqref{eq:vevA0} we get 
 the functional equation:
\begin{equation}
    \frac{x}{\phi(\mathfrak{q})} = \sum_{\nu} \mathfrak{q}^{|\nu|}   \frac{\prod_{\square \in \partial_+\nu }   Y(x + m c_\square)}{\prod_{\square \in \partial_-\nu }   Y(x + m c_\square)}
    \label{eq:qqchar-eq}
\end{equation}
for $Y(x)$, replacing the Eq. \eqref{ratcurve} of Vershik-Kerov problem. It would appear impossible to solve the infinite order non-linear difference equation Eq.\eqref{eq:qqchar-eq}. However it is solvable by what we call the \emph{$\theta$-transform}. Fix $z \in {\BC}^{\times}$, spectral parameter in the world of integrable systems. Apply
\begin{equation} \label{thetatransform}
    \chi(x)  \mapsto \sum_{n \in \mathbb{Z}} (-z)^n \mathfrak{q}^{\frac{n^2-n}{2}} \chi(x+ m n) \quad \text{for} \, z \in \mathbb{C}^*
\end{equation}
to both sides of the Eq. \eqref{eq:qqchar-eq}. Using factorization formula proven in the Appendix, one arrives at:
\begin{multline}
\label{eq:mastereq}
    x {\theta}(z; {\qe}) + m z \frac{d}{dz} \theta (z; {\qe}) 
    =\phi(\mathfrak{q}) Y(x) \times \\
    \prod_{n=0}^\infty \left(1 - z \mathfrak{q}^n \frac{Y(x+(n+1)m)}{Y(x+n m)} \right)  \prod_{n=1}^\infty \left(1 - z^{-1} \mathfrak{q}^n \frac{Y(x-n m)}{Y(x-(n-1) m)} \right)
\end{multline}
where:
\begin{gather}
   \theta(z ; {\qe}): \, = \, \sum_{n \in {\BZ}} (-z)^n \mathfrak{q}^{\frac{n^2-n}{2}} = 
   (1-z) \prod_{n=1}^{\infty} (1-{\qe}^{n}) (1 - z{\qe}^{n}) (1-z^{-1}{\qe}^{n}) 
\end{gather}
By comparing the expressions for zeros of the LHS and the RHS of the equality above we will be able to express $Y(x)$ in terms of $x$ as explicitly as needed to find the limit shape profile. Indeed, from the LHS we see that its zeros are described by (cf. \cite{NO} where this was derived by another method):
\begin{gather}
    x=x(z) = - m z \frac{d}{dz} \log \theta(z; {\qe}) = \\
    = m \left( \frac{z}{1-z} + \sum_{n=1}^{\infty} \left( \frac{z {\qe}^{n}}{1-z{\qe}^{n}} + \frac{{\qe}^{n}}{{\qe}^{n}-z} \right) \right)
\end{gather}
As function of $z$, $x$ obeys
\beq
x( {\qe}z) = x(z) + m 
\eeq
The real part of this function is depicted on figure (\ref{E1}) as a function of $\log z$.
Define the fundamental cylinder to be the annulus $|{\qe}|^{1/2}<|z|<|{\qe}|^{-1/2}$. On the fundamental cylinder the inverse function $z(x)$ is well defined. To first order in $\mathfrak{q}$ it looks like:
\begin{equation} \label{z(x)}
    z(x) = \frac{1}{1 + \frac{m}{x}} + O(\mathfrak{q})
\end{equation}
Hence we see that:
\begin{equation}
    z(x = \infty) = 1
\end{equation}
Now, let us have a look at the RHS of Eq. \eqref{eq:mastereq}. We see that the set of its zeros is a curve, whose branches are labeled by integers:
\begin{equation}
    z_n(x) \, = \, {\qe}^{n} \frac{Y(x-nm)}{Y(x+(1-n)m)}, \quad n \in \mathbb{Z}
\end{equation}
each behaving like:
\begin{equation}
    z_n(x) \rightarrow \mathfrak{q}^{n} \quad \text{as}\, \, x \rightarrow \infty
\end{equation}
In (\ref{z(x)}) we choose the $n=0$ branch:
\begin{equation}
    z(x) = z_0(x) = \frac{Y(x)}{Y(x+m)}
\end{equation}
Thus, we obtain the identity:
\begin{equation}
    \frac{d}{dx} \log z(x) = \frac{1}{2} \int_\mathbb{R} \frac{f''(y)}{x-y} dy - \frac{1}{2} \int_\mathbb{R} \frac{f''(y)}{x-y+m} dy 
\end{equation}
for the limit profile function $f \in C^{1}({\BR})$, which is related to $Y(x)$ as in \eqref{Ylimit}. From this we derive:
\begin{equation}
\frac{d}{dx} \log z(x) = \frac{1}{z\frac{dx}{dz}}   = \frac{1}{m F_{\mathfrak{q}}(z)}  
\end{equation}
where:
\begin{equation}
    F_{\mathfrak{q}}(z) = \sum_{n \in \mathbb{Z}} \frac{z \mathfrak{q}^n }{(1- z \mathfrak{q}^n)^2}
\end{equation}
one gets:
\begin{equation}
    \frac{1}{ F_{\mathfrak{q}}(z(x))}  = \frac{m}{2} \int_\mathbb{R} \frac{f''(y)}{x-y} dy - \frac{m}{2} \int_\mathbb{R} \frac{f''(y)}{x-y+m} dy 
    \label{eq:integraleq}
\end{equation}
The RHS of Eq. \eqref{eq:integraleq} has, as a function of $x$, two branch cuts a distance $m$ apart from each other, with the opposite sign jumps across each. As $x \to \infty$ the RHS of \eqref{eq:integraleq} goes to zero, as $(m/x)^2$, since 
\beq
\int_{\BR} f^{''}(y) dy = 2
\eeq
in agreement with the LHS.\\
The quasiperiodicity of $x(z)$ implies the branch cuts in the RHS correspond to the top and bottom edges 
of the fundamental cylinder in the $z$ variable:
$|z|= |{\qe}|^{1/2}$ and $|z|= |{\qe}|^{-1/2}$. Let us describe their locations explicitly.\\
For the bottom edge (parameterized by angle $\theta$) one has:
\beq \label{x}
    x(\mathfrak{q}^{1/2} e^{i \theta}) =  {\ii}  m \sin \theta \, g_{\mathfrak{q}}(\cos \theta)
\eeq
where:
\beq \label{g}
g_{\mathfrak{q}}(\cos \theta) = 2 \sum_{r \in \mathbb{Z}_{\geq 0} +\frac{1}{2}} \frac{\mathfrak{q}^r}{1 - 2 \mathfrak{q}^r \cos \theta + \mathfrak{q}^{2r}}    
\eeq
and for the upper edge we have:
\begin{equation}
    x(\mathfrak{q}^{-1/2} e^{\ii \theta}) =  x(\mathfrak{q}^{1/2} e^{\ii \theta}) - m
\end{equation}
As $X(\theta): = - {\ii} x(\mathfrak{q}^{1/2} e^{i \theta})/m$ is real for $\theta \in [- \pi, \pi]$ (it is depicted approximately for $\mathfrak{q} = 1/3$ on figure (\ref{E1real})) we see that one branch cut is located on a real axis, and another one is shifted from it into the imaginary direction by $-m$. Since $X({\theta})$ is odd:
\begin{figure}[h]
    \centering
    \includegraphics[scale = 0.8]{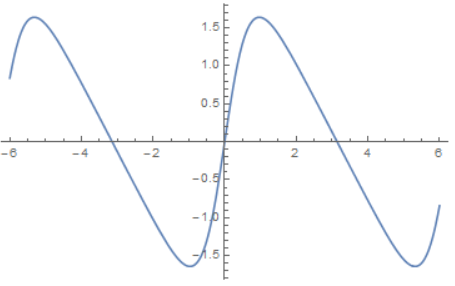}
    \caption{$X(\theta)$}
    \label{E1real}
\end{figure}
\begin{equation}
    X(- \theta) = - X(\theta) \, , 
\end{equation}
it vanishes at  ${\theta} = 0$ and ${\theta} = \pi$, therefore it has a maximum $\theta_* \in (0, \pi)$ :
\begin{equation}
    X'(\theta_* ) = 0 \ . 
\end{equation}
From the Eq. \eqref{eq:Xfrwp} below it follows it is unique. 
Accordingly,  $\pm x_*$, with
\begin{equation}
    x_*: = \ii m X(\theta_*)
\end{equation}
are the ends of the branch cut located on the real axis in the $x$-plane (see figure (\ref{mapgraph})). 
This means for any value of $x \in (- x_*, x_*)$ there are two corresponding values $\theta_{\pm}(x)$:
\begin{equation}
x =     {\ii} m X(\theta_+ (x)) = {\ii} m X(\theta_- (x))
\end{equation}
We denote the upper side of the cut by $C_+$, it is parametrized by $\theta$ running from $-\theta_*$ to $\theta_*$. We denote the lower side of the cut by $C_{-}$, it is parametrized by $\theta$ running from $-\pi$ to$ -\theta_*$, then from $\theta_*$ to $\pi$, see the figure (\ref{mappict}).
Therefore the jump which we are after is equal to:
\begin{equation}
\label{eq:fjump}
    \ii\pi m f''(x) = \frac{1}{F_{\mathfrak{q}} (\mathfrak{q}^{1/2} e^{\ii \theta_+})} - \frac{1}{F_{\mathfrak{q}} (\mathfrak{q}^{1/2} e^{\ii \theta_-})} .
\end{equation}
As
\beq
    X'(\theta) = F_{\mathfrak{q}} (\mathfrak{q}^{1/2} e^{i \theta}) = - \frac{\wp \Big(\frac{\tau}{2} + \frac{\theta}{2 \pi} \Big) }{4 \pi^2}    \, , 
    \label{eq:Xfrwp}
\eeq
the critical point $\theta_*$ is related to a zero of the Weierstrass function,. The latter has two zeroes on the elliptic curve, one giving the maximum of $X({\theta})$ for real $\theta$, and the other the minimum. 
The Eq. \eqref{eq:fjump} can be integrated once to give:
\begin{equation} \label{fmprime}
    f'(x) = \frac{\theta_+(x) - \theta_-(x)}{\pi} +1
\end{equation}
giving an elliptic version of arcsin law \cite{VK} of Vershik-Kerov. As the functions $\theta_+(x)$ and $\theta_-(x)$, so defined  that they are continuous on the interval $(-x_*, x_*)$ with $\theta_+(x_*) = \theta_-(x_*)$ obey $\theta_-(-x_*) - \theta_+(-x_*) = 2 \pi$, the choice of integration constant  ensures $f'(x_*) = - f'(-x_*) = 1$ (cf. Fig. \ref{mappict}).

\section{Limits and asymptotics}

\subsection{Edge asymptotics}
The functions $\theta_{\pm}(x)$ are transcendental, but the edge behavior of $f''(x)$ is easy to analyze: as $x \rightarrow \pm x_*$, $\theta \rightarrow \theta_*$, and we can expand
\begin{equation}
    X(\theta) = X(\theta_*) + \frac{1}{2} X''(\theta_*) (\theta - \theta_*)^2 + O \big((\theta - \theta_*)^3 \big)
\end{equation}
giving
\begin{equation}
    \theta_{\pm} = \theta_* \pm \sqrt{\frac{2(x - x_*)}{{\ii} m X''(\theta_*)}} \ ,  
\end{equation}
which by simple calculations leads to:
\beq
    f''(x) \sim \frac{2}{ \pi \sqrt{2 {\ii} m X''(\theta_*)(x-x_*)}} = \frac{\gamma}{\sqrt{x_* -x}}
    \label{eq:elledge}
\eeq
with 
\beq
{\gamma} = 2^{\frac 54} 3^{\frac 34} {\pi}^{-\frac 32} \left( 1 - 504 \sum_{n=1}^{\infty} n^5  \frac{{\qe}^{n}}{1-{\qe}^{n}} \right)^{-\frac 14}
\eeq
We can compare \eqref{eq:elledge} to \eqref{eq:limitVK}: 
\beq
f_{VK}^{''}(x) \sim \frac{1}{{\pi}\sqrt{\Lambda}} \frac{1}{\sqrt{2{\Lambda}-x}}
\eeq
Even though the functional forms of the edge asymptotics of the limit shapes in the elliptic and in the Vershik-Kerov case are similar, the detailed comparison requires a more precise matching of the parameters. We do this by taking the confluent (Inozemtsev) limit
\begin{multline}
    m \rightarrow \infty \, , \ 
    {\qe} \rightarrow 0 \, , \ 
    z \rightarrow 0\\
    {\qe}^{1/2} m = -{\ii} \Lambda \, \text{- fixed} \, , 
    m z = y \, \text{- fixed} 
\end{multline}
In this limit the measure in \eqref{partfunctmassive} reduces to  \eqref{eq:macro}. Next, the only terms left in the product from the RHS of \eqref{eq:mastereq} are:
\begin{equation}
    \Big(1 - \frac{y}{Y(x)}\Big) Y(x) \Big(1-y^{-1} \frac{{\Lambda}^2}{Y(x)} \Big)
\end{equation}
which is equal to:
\begin{equation}
 \chi(x) - \left( y + \frac{{\Lambda}^{2}}{y} \right)
\end{equation}
in agreement with the limit of \eqref{thetatransform}.
From \eqref{x} and \eqref{g} one sees that:
\begin{equation}
    x({\qe}^{1/2} e^{\ii \theta}) \rightarrow  2 \Lambda \sin \theta
\end{equation}
Hence, one has
\beq
   \theta_{+}(x) =  {\vartheta}_{+}^{VK}\left( \frac{x}{2 \Lambda} \right) := \arcsin \frac{x}{2 \Lambda} \, , \ 
    \theta_{+}(x) = {\vartheta}_{-}^{VK}\left( \frac{x}{2 \Lambda} \right) := \pi - \arcsin \frac{x}{2 \Lambda}
    \label{eq:vkangles}
\eeq
establishing the agreement between the Eqs. \eqref{fmprime} and \eqref{fprime}.\\
\subsection{Matching the edge behaviour.} The Inozemtsev limit of \eqref{eq:elledge}:
\begin{equation}
    f''(x) \sim  \frac{2}{ \pi \sqrt{2 {\ii}m X''(\theta_*)(x-x_*)}}  \rightarrow  \frac{1}{\pi \sqrt{\Lambda(2 \Lambda -x)}}
\end{equation}
matches the edge behaviour of $f_{VK}''(x)$:
\begin{equation}
    f_{VK}''(x) = \frac{1}{ \pi \Lambda } \frac{1}{\sqrt{1-\big(\frac{x}{2\Lambda}\big)^2  }} \sim \frac{1}{\pi \sqrt{\Lambda(2 \Lambda -x)}}
\end{equation}

\subsection{Expanding around Vershik-Kerov limir shape}
\begin{figure}[t]
    \centering
    \includegraphics[scale = 0.8]{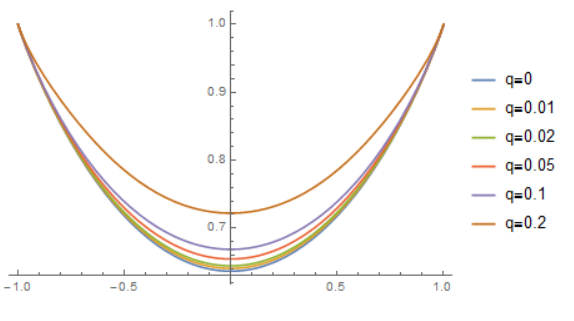}
    \caption{$f(x)$ to the first few orders in $\mathfrak{q}$}
    \label{fm}
\end{figure}

The comparison of the limit shape of our problem to that of \eqref{eq:macro} is an instructive exercise in perturbative renormalization. Naively, fixing $\Lambda = {\ii}m {\qe}^{\frac 12}$ and varying $\qe$, 
for small $\qe$  we can find $\theta_{*}$, ${\theta}_{\pm}(x)$ by 
expanding in ${\qe}^{\frac 12}$, cf. \eqref{eq:vkangles}:
\begin{multline}
    {\theta}_{*} = \frac{\pi}{2} -  {\qe}^{\frac 12} \left( 2 -  \frac 83 {\qe} + \frac{72}{5}{\qe}^{2}-\frac{632}{7}{\qe}^{3} + \frac{5462}{9} {\qe}^{4} - \frac{47016}{11} {\qe}^{5} + \ldots \right) \, , \\
    x_{*} = 2{\Lambda} \left( 1 + 2{\qe} + 8 {\qe}^{3} - 29 {\qe}^{4} + 162 {\qe}^{5} + \ldots \right) \, , 
    \label{eq:qexp}\end{multline}
    which then leads to the naively singular
    expanstion for $\theta_{\pm}(x)$ and $f(x)$:
    \begin{multline}
    {\theta}_{\pm}(x) \, =^{\kern -.1in?} \ {\vartheta}_{\pm}^{VK}\left( {\xi} \right) - 2 {\qe}^{\frac 12} {\xi}  \mp  2{\qe}  \frac{{\xi}^{3}}{\sqrt{1-{\xi}^{2}}} + 4 {\qe}^{\frac 32} {\xi} \left( 1 + \frac{2}{3} {\xi}^{2} \right)  + O({\qe}^{2}) \, , 
\\
 f'(x) \, =^{\kern -.1in?} \  \frac{2}{\pi}  \arcsin {\xi} - \, \frac{4{\qe}}{\pi}   \frac{{\xi}^3}{\sqrt{1- {\xi}^2}} + O({\qe}^2) \, , \\
\end{multline}
with $\xi = x/2{\Lambda}$. There are, 
of course,  no singular terms in $f^{\prime}(x)$, as it is a monotone continuous function
on $[-x_{*}, x_{*}]$, changing from $-1$ to $+1$. The resolution of the puzzle is that the singularities reflect the $\qe$-dependence of the cut. If 
instead of $\Lambda$ one keeps fixed $x_{*}$, the corresponding expansion becomes perfectly non-singular:
\begin{multline}
 {\theta}_{\pm}(x) \, = \,  
 {\vartheta}_{\pm}^{VK}\left( y\right) - 
 2 {\qe}^{\frac 12} y \, \times \\
 {\scriptscriptstyle{\left( 1 - \frac{4{\qe}}{3} y^{2}
 + 4 {\qe}^{2} \left( 1 - 2 y^{2}  - \frac{4}{5} y^{4} \right) -  20 {\qe}^{3} \left( 1 + \frac{2}{5} y^{2}
 - \frac{16}{5} y^{4} - \frac{16}{35} y^{6} \right) + 61 {\qe}^{4} 
 \left( 1 + \frac{272}{61} y^{2} - \frac{224}{61} y^{4} - \frac{384}{61} y^{6} - 
 \frac{256}{549} y^{8} \right) + \ldots \right)}} + \\
 \mp  2{\qe} y \sqrt{1-y^{2}} \, \times \\
 {\scriptscriptstyle{\left(  1 - 3 {\qe} (1 +\frac 23 y^2) + 6{\qe}^{2} \left( 
 1+ \frac{40}{9} y^{2} + \frac{8}{9} y^{4} \right) - 3 {\qe}^{3} \left(  1 + 
 38 y^2 + 56 y^4 + \frac{16}{3} y^{6} \right) +
 \frac{7296}{5} {\qe}^{4} y^{4} \left( 1 + \frac{12}{19}y^{2} + \frac{2}{57} y^4 \right) + \ldots \right)}}  \, ,  
\label{eq:qexpimpr}\end{multline}
where $y =x/x_{*}$. For the values of $\mathfrak{q} = 0, \, 0.01, \, 0.02, \, 0.05, \, 0.1, \, 0.2$ and $x_* = 1$ the
plot of function $f(x)$ is drawn in the figure (\ref{fm}).

\section{Conclusions and future directions}

In this paper we explored a one-parametric deformation ${\bmu}_{m, {\qe}, {\hbar}}$ of the Plancherel measure on the set of Young diagrams, and found 
its limit shape. Specifically, we kept the size fugacity $\qe$ finite, and tuned the parameter
$m/{\hbar} \to \infty$.

There is a natural generalization of such limit shape problem (motivated, e.g. by topological string theory and gauge theory \cite{OP,LMN}), where the measure ${\mu} [{\lambda}]$ is  includes the chemical potentials for the generalized Casimirs ${\bf p}_{k}$. 
Introduce the sequence $(t_k)$, $k = 1, 2, \ldots$ of formal variables, the formal function
\beq
{\bf t}(x) = \sum_{k=1}^{\infty} t_{k}x^{k}
\eeq
and define the measures (cf. \cite{LMN})
\beq
\begin{aligned}
& {\mu}_{\Lambda, \hbar; \bf t} [{\lambda}] = {\mu}_{\Lambda, \hbar} [{\lambda}] \,   e^{{\bf t}({\hbar}c_{\square})} \\
& {\mu}_{m, {\qe}, \hbar; \bf t} [{\lambda}] = {\mu}_{m , {\qe}, \hbar} [{\lambda}] \,   e^{{\bf t}({\hbar}c_{\square})} \end{aligned}
\label{eq:hightimes}
\eeq
The limit shape is now governed by the analytic multi-valued function $Y(x)$, which behaves as $x + o(x^{-1})$
on the physical sheet, such that, in the first case, 
\beq
Y(x) + {\Lambda}^{2} \frac{e^{{\bf t}(x)}}{Y(x)} 
\label{eq:chart}
\eeq
is an entire function of $x$. One cannot claim \eqref{eq:chart} is a linear function of $x$, as the LHS has an essential singularity at $x \to \infty$. However, the formal nature of variables $t_k$ suggests
the Riemann surface of $Y$ is still a two-sheeted cover of the $x$-plane. Also, the probabilistic nature of the problem shows $Y$ has a single cut on the physical sheet. By comparing the two terms in the LHS of \eqref{eq:chart}
one concludes
$Y$ is an analytic function on the curve $\CalC$
\beq
y + \frac{{\tilde\Lambda}^{2}}{y} = x + {\tilde v}
\eeq
where the parameters $({\tilde v}, {\tilde \Lambda})$ are the formal functions of $t_k$, such that at
$t_k =0$ they approach $(0, {\Lambda})$. There are two special points $P_{\pm}$, where $x = \infty$. 
At $P_+$, $y \sim x$ and at $P_{-}$, $y \sim {\Lambda}^{2} x^{-1} \to 0$. In other words
$(x,y) = ({\infty}, {\infty})$ at $P_{+}$, and $(x,y) = ({\infty}, 0)$ at $P_-$. 
The function $Y$ is found from the following conditions: it is holomorphic on $\CalC$ outside $P_{\pm}$, and it has the asymptotics
\beq
\begin{aligned}
 Y \sim x \, ,& \ (x,y) \to P_{+} \, , \\
Y \sim {\Lambda}^{2} x^{-1} e^{{\bf t}(x)} \, , &\ (x,y) \to P_{-} 
\end{aligned}
\label{eq:Yasympt}
\eeq
Here is how one finds such a function: Define the functions ${\Omega}_{k}^{\pm}$
to be meromorphic functions on $\CalC$, holomorphic outside $P_{\pm}$, respectively, such that
\beq
{\Omega}_k^{\pm} = x^{k} + o(x^{-1}) \, , \ (x,y) \to P_{\pm} \ .
\eeq
It follows
\beq
x^{k} = {\Omega}_{k}^{+}(y) + {\Omega}^{-}_{k}(y) - {\omega}_{k} \, , 
\eeq
where
\beq
{\omega}_{k} = {\Omega}_{k}^{+}(P_{-}) = {\Omega}_{k}^{-}(P_{+}) = \frac{1}{2\pi\ii} \oint_{|y|= |{\tilde\Lambda}|} \frac{dy}{y} x^{k}\ .
\eeq
Then
\beq
Y = y \, {\exp}\, \sum_{k} t_{k} ( {\Omega}^{-}_{k}(y)-{\omega}_{k}) 
\eeq
has the correct asymptotics both at $P_{\pm}$, and continues analytically across the cuts of the $y$-function, provided that the matching equations at the branch-points 
\beq
 {\tilde\Lambda}^2 e^{-\sum_{k} t_{k}  {\omega}_{k}} = {\Lambda}^{2}
 \eeq
hold and the vanishing period
\beq
0 = - \oint x \frac{dY}{Y} \, = \, {\tilde v} +  \sum_{k} \, t_{k}\,  {\rm Coeff}_{y^{-1}} \, {\Omega}_{k}^{-} 
\eeq
guarantees the correct asymptotics $Y \sim x + o(x^{-1})$ on the physical sheet. 
This gives two equations for the two unknowns $({\tilde v}, {\tilde \Lambda})$. For example, setting $t_{2} = t_{3} = \ldots = 0$ one easily recovers the Lambert solution found in \cite{MN}:
\beq
{\tilde v} = - t_{1} {\tilde\Lambda}^{2}\, , \ {\Lambda}^{2} = {\tilde \Lambda}^{2} e^{-2t_{1} {\tilde\Lambda}^{2}}
\eeq
with its finite convergence radius typical of Whitham hierarchies \cite{KW}.  The second case of \eqref{eq:hightimes}
and other generalizations will be presented in the companion paper \cite{GN2}.

There is yet another class of limit shape problems, where the parameter $m/{\hbar}$ in ${\bmu}_{m, {\qe}, {\hbar}}$ is fixed while the instanton fugacity $\qe$ approaches  $1$, a Hardy-Ramanujan limit. In the special case of $m=0$, the limit shape curve is the celebrated
\beq
e^{-a} + e^{-b} = 1 \, , \ a, b \geq 0
\eeq
The generalizations to $m \neq 0$ will be considered elsewhere.

Finally, all the measures discussed above were symmetric under ${\lambda} \mapsto 
{\lambda}^{t}$. Four dimensional gauge theory \cite{Nekrasov:2002qd} suggests
yet another natural generalization, in which the weight of the box 
$\square \in \lambda$ depends separately on the arm $a_{\square} = {\lambda}_{i} - j$ and the leg $l_{\square} = {\lambda}_{j}^{t} - i$, for example
\begin{multline}
{\bmu}_{m, {\qe}; {\ve}_{1}, {\ve}_{2}} [ {\lambda} ] = 
\frac{1}{Z_{2^{*}}(m, {\qe}; {\ve}_{1}, {\ve}_{2})} \ {\qe}^{|{\lambda}|} \ \times \\
\prod_{\square} \left( 1  + \frac{m}{{\ve}_{1} (a_{\square}+1) - {\ve}_{2} l_{\square}} \right) 
\left( 1  + \frac{m}{-{\ve}_{1} a_{\square}+ {\ve}_{2} (l_{\square}+1)} \right) 
\end{multline}
Of course, such measures are well-known to mathematicians under the name of discrete $\beta$-ensembles, Jack processes, etc.  \cite{BO,DK}. 
We know   
\cite{BPSCFT1} the partition function exponentiates 
\beq
Z_{2^{*}}(m, {\qe}; {\ve}_{1}, {\ve}_{2}) \sim {\exp} \frac{1}{\ve_2} W ( m , {\qe}; {\ve}_{1})
\eeq 
when $\ve_2 \to 0$ with $m, {\ve}_{1}, {\qe}$ kept constant. 
The corresponding limit shape is described by a quantum spectral curve \cite{NG}.

{\bf Acknowledgements.} We have greatly benefited from patient explanations of I.~Krichever and A.~Okounkov. Research is partly supported by NSF PHY Award 2310279. 

\section{Appendix. Proof of the factorization formula}
The $\hbar \to 0$ limit of the normalized $qq$-character, including the higher times,  is given by
\begin{equation} 
    {\chi}(x) = {\phi}({\qe}) \sum_{\lambda} \mathfrak{q}^{|\lambda|} \prod_{\square \in \lambda} e^{t(x + m c_{\square})}  \frac{\prod_{\square \in \partial_+\lambda }   Y(x + m c_\square)}{\prod_{\square \in \partial_-\lambda }   Y(x + m c_\square)}
\end{equation}
Motivated by \cite{NG} we prove the following
$\mathsf{Lemma}$: Let $z$ be an indeterminate. The following identity  holds:
\begin{multline} \label{factorizationform}
    Y(x) \cdot
    \prod_{n=0}^\infty \left(1 - z \mathfrak{q}^n e^{ \hat{t}(x+m n) } \frac{Y(x+(n+1)m)}{Y(x+n m)} \right)  \times \\ \prod_{n=1}^\infty \left(1 - z^{-1} \mathfrak{q}^n e^{ -\hat{t}(x-m n) } \frac{Y(x-n m)}{Y(x-(n-1) m)} \right) = \\
\sum_{n=1}^\infty (-z)^n \mathfrak{q}^{\frac{n^2-n}{2}} e^{\sum_{j=0}^{n-1} \hat{t}(x+j m)} \chi(x + n m) + \chi(x) + \\
+ \sum_{n=1}^\infty (-z)^{-n} \mathfrak{q}^{\frac{n^2+n}{2}} e^{-\sum_{j=1}^{n} \hat{t}(x-j m)} \chi(x - n m)
\end{multline}
where $\hat{t}(x)$ a unique formal power series in $x$, ${\hat t}(0) = 0$, solving:
\begin{equation}
    t(x) = \hat{t}(x) - \hat{t}(x-m)
\end{equation}
\textbf{Proof}. $\square$
By a simple cancellation of factors the formula for the $qq$-character could be rewritten as:
\begin{equation} \label{qChar}
    \chi(x) =  \sum_{\lambda}  \mathfrak{q}^{|\lambda|} \prod_{\square \in \lambda} e^{t(x + m c_{\square})}  \prod_{j =1}^{\lambda_1}  \frac{Y(x + m(\lambda_j^t-j+1) ) }{Y(x + m(\lambda_j^t-j)  )} Y (x - m \lambda_1) 
\end{equation}
 Opening the brackets in the LHS of the formula (\ref{factorizationform}) one obtains:
 \begin{multline}
     \sum_{r , s\geq 0} \sum_{\substack{ n_0 > n_1 >...>n_{r-1} \geq 1 \\ 0 \leq k_0 < k_1<...< k_{s-1}}} (-z)^{r-s} \prod_{i=0}^{r-1} \mathfrak{q}^{n_i-1} e^{\hat{t}(x + (n_i-1)m)} \prod_{i=0}^{s-1} \mathfrak{q}^{k_i+1} e^{-\hat{t}(x - (k_i+1)m)} \\
     \cdot 
\prod_{i =0 }^{r-1} \frac{Y (x+ n_i m  ) }{Y (x+ (n_i -1 )m  )} \cdot Y(x) \cdot \prod_{i = 0}^{s-1} \frac{Y (x - (k_i +1)m }{Y(x - k_i m )} 
 \end{multline}
Note that the two sets of strictly increasing numbers $n_0 > n_1 >...>n_{r-1} \geq 1$ and $0 \leq k_0 < k_1<...< k_{s-1}$ encode the information about a Young diagram $\lambda$ and an additional integer, which could be interpreted as a shift of the Young diagram perpendicular to the main diagonal. The dictionary is the following. The shift is equal to $p = r-s$. The positive integers $n_j$ define the lengths of the first $r$-columns, and $k_i$ define the length of the first $s$-rows, through the formulas:
 \begin{gather}
     n_j = \lambda_{j+1}^t -j +p, \qquad j = 0,...,r-1 \\
     k_{s-i} = \lambda_i - i - p , \qquad i = 1,...,s
 \end{gather}
 This data uniquely determines the diagram. Notice that, given $\lambda$ and $p$ the numbers $r$ and $s$ are uniquely determined as such values of $i$ and $j$
 where the expressions $\lambda_i - i - p $, $\lambda_{j+1}^t -j +p$ change sign.
 With this substitution the above expression could be rewritten as follows:
 \begin{multline}
     \sum_{p \in \mathbb{Z} } \sum_{\lambda} (-z)^{p} \prod_{j=0}^{r-1} \mathfrak{q}^{\lambda_{j+1}^t -j +p-1} e^{\hat{t}(x+ m(\lambda_{j+1}^t -j +p-1) )} \times \\
     \prod_{i=1}^{s} \mathfrak{q}^{\lambda_i -i -p +1} e^{\hat{t}(x+ m(\lambda_i -i -p +1) )}\, \times Y(x)  \times  \\
\prod_{j =0 }^{r-1} \frac{Y (x+ (\lambda_{j+1}^t -j +p) m ) }{Y (x+ (\lambda_{j+1}^t -j +p -1 )m )} 
\times \prod_{i = 1}^{s} \frac{Y (x - (\lambda_i - i - p+1)m )}{Y (x - (\lambda_i - i - p) m)} 
 \end{multline} 
 Now we need to match every multiple in the product to every multiple in the expression \eqref{qChar}, shifted by $p$.  Let us denote:
 \begin{multline}
 \text{LHS}(\lambda,p)  = \prod_{j =0 }^{r-1} \frac{Y (x+ (\lambda_{j+1}^t -j +p) m ) }{Y (x+ (\lambda_{j+1}^t -j +p -1 )m )} \cdot Y(x) \cdot 
\prod_{i = 1}^{s} \frac{Y (x - (\lambda_i - i - p+1)m )}{Y (x - (\lambda_i - i - p) m)}   
 \\
 \text{RHS}(\lambda,p) = Y (x - m (\lambda_1 - p) )  \prod_{\substack{j =1 } }^{\lambda_1} \frac{Y(x + m(\lambda_j^t-j+1+p)  }{Y(x + m(\lambda_j^t-j+p) )}     
 \end{multline}
 We are going to prove that $\text{LHS}(\lambda,p) = \text{RHS}(\lambda,p)$ by induction on the number of boxes in the Young diagram.\\
 The base of the induction is the case when $\lambda = \varnothing$,  and either $r = 0$, and hence $s = -p$, or $s=0$, and $r=p$.\\
 Let us consider the case $r = 0$ first.
The LHS$(\varnothing,-s)$ of the formula above then takes the form:
\begin{equation}
Y(x)
\cdot \prod_{i = 1}^{s} \frac{Y (x +(i + p-1)m)}{Y (x +(i + p)m)}    
\end{equation}
which is, after cancelling all factors, is equal to the RHS$(\varnothing,-s) = 
Y(x + m p ) $. Letting now $s = 0$, one has:
\begin{multline}
\text{LHS}(\varnothing,r) = \\ = \prod_{j =0 }^{r-1} \frac{Y (x+ (r-j) m ) }{Y (x+ (r -j -1 )m )} Y(x-(r-1) )  =
Y(x + r m)
\end{multline}
which is equal to the RHS$(\varnothing,r)$.\\
For the induction step, let us assume, that we are adding one box to the $k$'th row. For the LHS the cases $k - \lambda_k -1 +p \geq 0$ and $k - \lambda_k -1 +p < 0$ should be treated separately, because they affect the product of the first $r$ factors or the last $s$ factors correspondingly, but eventually the final result is the same: 
\begin{multline}
    \frac{\text{LHS}(\lambda+1_k,p)}{\text{LHS}(\lambda,p)} = \\ =\frac{Y(x-(\lambda_k -k -p +2)m )}{Y(x-(\lambda_k -k -p +1)m )} \frac{Y(x-(\lambda_k -k -p)m )}{Y(x-(\lambda_k -k -p +1)m)}
\end{multline}
By analogous calculation same is for the RHS. Hence the formula is proven.\\
Now let us deal with the factors depending on $\hat{t}(x)$. In the LHS we have the function in the exponent:
\begin{equation}
   d(\lambda,p): =  \sum_{i=0}^{r-1} \hat{t}(x + m(n_i-1)) - \sum_{j=0}^{s-1} \hat{t}(x - m(k_j+1))
\end{equation}
Similarly to the discussion above we could calculate its change under the addition of the box into the $k$'s row:
\begin{equation}
   d(\lambda+ 1_k,p) -  d(\lambda,p) = \hat{t}(x - m (\lambda_k - k -p +1)) - \hat{t}(x - m (\lambda_k - k -p +2))
\end{equation}
For the RHS we would like to look at the function:
\begin{multline}
    d'(\lambda,p): = \sum_{\square \in \lambda} t(x + m (p+c_\square)) = \sum_{\square \in \lambda} \hat{t}(x + m(p+ c_\square)) - \hat{t}(x + m(p-1+ c_\square))
\end{multline}
And hence:
\begin{equation}
   d'(\lambda+ 1_k,p) -  d'(\lambda,p) = \hat{t}(x - m (\lambda_k - k -p +1)) - \hat{t}(x - m (\lambda_k - k -p +2))
\end{equation}
So the step of the induction is proven. And for the base we have:
\begin{equation}
    d'(\varnothing, p) = 0
\end{equation}
However:
\begin{equation}
    d(\varnothing, p) = \begin{cases}
        \sum_{j=0}^{p-1} \hat{t}(x+j m), \quad p \geq 0 \\
        -\sum_{j=1}^{-p} \hat{t}(x-j m), \quad p < 0
    \end{cases}
\end{equation}
exactly the factors we see in (\ref{factorizationform}). The last step is to compare the $\mathfrak{q}$ dependence. This proof is carried out by the same trick. 

Note that the proof is similar to fermionic proof of Jacobi triple product identity.
$\blacksquare$


\begin{thebibliography}{}

\bibitem{BO} A.~Borodin, G.~Olshanski, \emph{$z$-measures on partitions and their scaling limits},  arXiv e-prints. doi:10.48550/arXiv.math-ph/0210048

\bibitem{DK} E.~Dimitrov,  and A.~ Knizel, \emph{Asymptotics of discrete $\beta$-corners processes via two-level discrete loop equations},  Probability and Mathematical Physics 3.2 (2022): 247-342.

\bibitem{DW}
R.~Donagi and E.~Witten,
\emph{Supersymmetric Yang-Mills theory and integrable systems},
Nucl. Phys. B \textbf{460}, 299-334 (1996)
doi:10.1016/0550-3213(95)00609-5
[arXiv:hep-th/9510101 [hep-th]].

\bibitem{NG} A.~Grekov, N.~Nekrasov, \emph{Elliptic Calogero-Moser system, crossed and folded instantons, and bilinear identities}, (2023)	arXiv:2310.04571 [math-ph] 

\bibitem{GN2} A.~Grekov, N.~Nekrasov, \emph{Vershik-Kerov in higher times and higher spaces}, to appear.



\bibitem{VK}  S.~Kerov and A.~Vershik, \emph{Asymptotics of the Plancherel measure of the symmetric group and the limiting form of Young tableaux}, Doklady akademii nauk, Russian Academy of Sciences  Vol. {\bf 233} (1977) 6, pp. 1024-1027, MR0480398 


\bibitem{KW}
I.~Krichever,
\emph{The $\tau$-function of the universal Whitham hierarchy, matrix models and topological field theories},
Commun. Pure Appl. Math. \textbf{47}, 437 (1994)
[arXiv:hep-th/9205110 [hep-th]].


\bibitem{LS} B.~Logan and L.~Shepp, \emph{A variational problem for random Young tableaux}, Adv.Math. {\bf 26} (1977) 206-222, MR1417317

\bibitem{LMN}
A.~S.~Losev, A.~Marshakov and N.~A.~Nekrasov,
\emph{Small instantons, little strings and free fermions},
[arXiv:hep-th/0302191 [hep-th]].

\bibitem{MN}
A.~Marshakov and N.~Nekrasov,
\emph{Extended Seiberg-Witten Theory and Integrable Hierarchy},
JHEP \textbf{01}, 104 (2007)
doi:10.1088/1126-6708/2007/01/104
[arXiv:hep-th/0612019 [hep-th]].

\bibitem{NO} N.~Nekrasov and A.~Okounkov, \emph{Seiberg-Witten theory and random partitions}, In,  `The Unity of Mathematics: In Honor of the Ninetieth Birthday of I.M.~Gelfand', Boston, MA: Birkh{\"a}user Boston (2006) 525-596.

\bibitem{Nekrasov:2002qd}
N.~A.~Nekrasov,
\emph{Seiberg-Witten prepotential from instanton counting},
Adv. Theor. Math. Phys. \textbf{7}, no.5, 831-864 (2003)
doi:10.4310/ATMP.2003.v7.n5.a4
[arXiv:hep-th/0206161 [hep-th]].
\bibitem{BPSCFT1} N.~Nekrasov, \emph{BPS/CFT correspondence: non-perturbative Dyson-Schwinger equations and qq-characters}, Journal of High Energy Physics,  {\bf 3} (2016)  1-70 
\bibitem{NP} N.~Nekrasov, V.~Pestun, \emph{Seiberg-Witten Geometry of Four-Dimensional ${\CalN}=2$ Quiver Gauge Theories}, arXiv:1211.2240v2 [hep-th]




\bibitem{OP} A.~Okounkov,  and R.~Pandharipande, \emph{Gromov-Witten theory, Hurwitz theory, and completed cycles},
(2002) arXiv:0204305  [math.AG]

\bibitem{VW}
C.~Vafa and E.~Witten,
\emph{A Strong coupling test of S duality},
Nucl. Phys. B \textbf{431}, 3-77 (1994)
doi:10.1016/0550-3213(94)90097-3


\end{thebibliography}
\end{document}